\documentclass[12pt]{iopart}
\usepackage{amsmath}
\usepackage{iopams}

\begin{document}

\maketitle

\title{Null Lagrangians and Gauge Functions in Dynamical Systems: 
Forces and Nonlinearities}

\author{L.C. Vestal and Z.E. Musielak}
\address{Department of Physics, University of Texas 
at Arlington, Arlington, TX 76019, USA \\}
\ead{zmusielak@uta.edu}

\begin{abstract}
Among different Lagrangians, null Lagrangians are known for having 
identically zero the Euler-Lagrange equation and, therefore, they have no effects 
on the resulting equations of motion.  However, there is a special family of null 
Lagrangians that can be used to convert linear and undriven equations of motion 
into nonlinear and driven ones.  To identify this special family, general null 
Lagrangians and their gauge functions are constructed for second-order ordinary 
differential equations of motion describing one-dimensional dynamical systems.
The gauge functions corresponding to forces and nonlinearities in a variety of 
known dynamical systems are presented and novel roles of these functions in 
dynamics are discussed.
\end{abstract}

\section{Introduction}

In modern physics, the equations of motion for different physical systems 
are obtained by using the laws of physics.  For a given physical system, 
the equation of motion can also be derived by using the principle of least 
action, whose Lagrangian, when substituted into the Euler-Lagrange (EL),
equation gives the required equation.  The procedure for deriving the 
equation of motion from a given Lagrangian is effective and elegant.
However, the problem is that the Lagrangian is typically found after the 
equation of motion is already known because the first principle theory  
to derive Lagrangians for different equations of physics is not currently 
known. 

Among different Lagrangians that are known, the so-called null (or trivial) 
Lagrangians (NLs) play a special role, as they have identically zero the 
EL equation [1]. Another important property of the NLs is that they can be 
expressed as the total derivative of a scalar function [1], which we refer to as 
a gauge function. The existence of different Lagrangians is guaranteed by the 
Helmholtz conditions [2,3], except NLs, whose existence is independent from 
these conditions.  In mathematics, NLs have been investigated for many years; 
the studies have involved structures and constructions of these Lagrangians 
[4,5], their geometric formulations [6], Cartan and Lepage forms and 
symmetries of Lagrangians [7,8], and their roles in field theories [9,10], 
specifically in Carath\'eodory's theory of fields of extremals and integral 
invariants [5,11].  However, there are only several known studies of NLs 
in physics, with the most prominent being applications to elasticity, where 
they represent the energy density function of materials [12,13], and more 
recently to classical mechanics, where they are used to demonstrate the 
Galilean invariance of the Lagrangian for the Newtonian law of inertia 
[14,15] and to introduce forces [16-18].

Methods to construct NLs were described previously in the literature
[1,4-10].  Recently, a different approach was considered by restricting the
orders of the dependent and independent variables.  The constructed NLs
are of lower or comparable orders to the Lagrangian for a given equation 
and depend on arbitrary constant coefficients [14].  This method was 
generalized by replacing the constants with arbitrary functions of the 
independent variable [15-20].  Moreover, the concept of exact NLs was 
introduced based on the invariance of the action and it was shown that, 
after having obtained the exact NLs, their corresponding exact gauge 
functions can be derived [15].  Some of these NLs were used to 
introduce forces to the law of inertia and converting it into the second 
law of dynamics [16].  It was shown that some simple NLs can be used 
to add time-dependent forces to a harmonic oscillator [17] and to the 
Bateman oscillator [18], and converting these undriven systems into 
the driven ones.  Our previous work is significantly generalized in this
paper, which demonstrates how to construct general gauge function 
and their corresponding NLs, and used them to convert undriven 
dynamical systems into the driven ones, and linear systems into 
the nonlienar ones.

The main objective of this paper is to construct the gauge functions
and their corresponding NLs for second-order ordinary differential 
equations (ODEs) in one dimension, and apply the obtained results to 
a variety of dynamical systems.  It must be pointed out that the forms 
of NLs for second-order ODEs are known [1,4-6] and that the NLs 
constructed in this paper are consistent with them, and in special 
cases they reduce to the NLs found in [16,17]. Nevertheless, the 
main purpose of the NLs constructed here is to use them to introduce 
forces and nonlinearities into otherwise undriven and linear dynamical
systems, which has not yet been done.  This is achieved by using the 
Legendre transform to generate new energy terms and adding them 
to the standard Lagrangian of a dynamical system, which modifies 
the original system; this is the main novelty of our approach described
in this paper.  We explore physical effects of such modifications 
and demonstrate that forces and nonlinearities can be introduced 
to dynamical systems this way.  The presented gauge functions for 
a variety of known dynamical systems are our main results as these 
functions can be used to determine the corresponding forces and 
nonlinearities that are introduced into otherwise undriven and linear 
dynamical systems.

Our paper is structured as follows: our method to construct general null 
Lagrangians analytically is described in Section 2; applications of the 
obtained results are discussed in Section 3; and our conclusions are 
given in Section 4. 

\section{Construction of null Lagrangians}
\subsection{Null Lagrangians and gauge functions}

Let $L_s$ be a standard Lagrangian, whose kinetic and potential energy 
terms can be easily identified, and whose substitution into the E-L equation 
gives a second-order ODE.  Let $x (t)$ be the dependent variable of this
ODE and $t$ be the independent variable.  Then, the standard Lagrangian 
must depend on $\dot x$, which is the first derivative of $x$ with respect 
to $t$, and also on $x$, so we may write $L_s (\dot x, x)$, with $x$ and 
$\dot x$ being ordinary ($x :\ \mathcal{R} \rightarrow \mathcal{R}$) 
and smooth ($\mathcal{C}^{\infty}$) functions.  In general, the standard 
Lagrangian may also depend explicitly on $t$, thus, $L_s (\dot x, x, t)$,
which means that this Lagrangian describes damped and amplified oscillators 
[18], such as Bateman oscillators [19] for which $L_s (\dot x, x, t)$ becomes 
the Caldirola-Kanai Lagrangian [20,21]. 

If $L_{n}$ is a NL and $\hat {EL}$ is the Euler-Lagrange 
operator, then $\hat {EL} (L_{n}) = 0$, as required by the definition of 
the NLs.  Such NLs may depend on both the dependent and independent 
variables, so they can be written as $L_{n} (\dot x, x, t)$.  These NLs can be 
added to the standard Lagragian, $L_s (\dot x, x, t)$, without having any 
effect on the resulting equation of motion.  Thus, the total Lagrangian, 
$L_{tot}$, can be written as 
\begin{equation}
L_{tot} (\dot x, x, t) = L_s (\dot x, x, t) + L_{n} (\dot x, x, t)\ ,
\label{eq1}
\end{equation}
where $L_{n} (\dot x, x, t)$ is the NL that can be obtained 
by calculating the total derivative of any scalar and differentiable function [1,4-6],
which is called the gauge function [14-17] and denoted as $\Phi (x, t)$, with $x 
= x (t)$.  Thus, the general NL can be written as  
\begin{equation}
L_{n} (\dot x, x, t) = {{d \Phi} \over {dt}} = {{\partial \Phi} \over 
{\partial t}} + \dot x {{\partial \Phi} \over {\partial x}}\ .
\label{eq2}
\end{equation}
Substitution of this NL into the EL equation shows that the gauge function 
can be of any form as long as it has the properties specified above.  

Despite the fact that the presence of NLs does not change the form of 
the resulting equation of motion, it has been shown that the NLs may be 
used to restore Galilean invariance of standard Lagrangians [15], and to 
independently introduce forces to classical mechanics by some, but not all, 
NLs [16-18].  Therefore, it is desired to construct general NLs and identify 
among them those NLs that can be used to introduce forces; it must be 
pointed out that the NLs for second-order ODEs are known [1,4-6] and 
that the NLs constructed in this paper are consistent with those previously 
obtained.  

The previous work on defining forces [16-18] is extended in this paper 
to specific forces that frequently appear in dynamics as well as to nonlinearities 
that are present in some well-known dynamical systems.  For the considered
forces and nonlinearities their gauge functions are given.  These gauge 
functions can be used to obtain the corresponding NLs and to convert 
undriven and linear dynamical systems into the driven and nonlinear 
systems.  In the following, the procedure that allows for the conversion, 
and uses the Legendre transform to generate new energy terms that are 
then added to the standard Lagrangian of a given dynamical system, is
described, and then applied to some well-known oscillators in classical 
dynamics.  

\subsection{Energy function and definition of forces}

Let us consider a conservative dynamical system, whose equation of motion 
is a second-order differential equation that can be obtained from the standard
Lagrangian,  $L_s (\dot x, x)$, that does not depend explicitly on time.
Moreover, let us assume that the NL resulting from Eq. (\ref{eq2}) depends 
explicitly on time.   Then, the energy function must be calculated [22,23],
and, for the total Lagrangian, $L_{tot} (\dot x, x, t) = L_s [\dot x, 
x] + L_{n} (\dot x, x, t)$, the energy function is given by 
\begin{equation}
E_{tot} (\dot x, x, t) = \dot x {{\partial (L_{s} + L_{n})} \over 
{\partial \dot x}} - (L_{s} + L_{n}) = E_{s} (\dot x, x) + 
E_{n} (x, t)\ .
\label{eq3}
\end{equation}
where $E_{s} (\dot x, x)$ represents the total energy of the system,
which is equal to its Hamiltonian that can be used to derive the equation
of motion from the Hamilton equations [22,23].  The Legendre 
transformation relates the Lagrangian and Hamiltonian formulations
[24], and leads to the same Hamilton equations independently from
the null Lagrangian that is added to the standard Lagrangian.  The 
necessary condition is $dE_{tot} / dt = - [\partial (L_{s} + L_{n}) / 
\partial t]$, which plays the same role as the EL equation does for
$L_{s}(\dot{x})$ [22-24].

In addition, $E_{n} (x, t)$ becomes 
\begin{equation}
E_{n} (x, t) = - {{\partial \Phi} \over {\partial t}}\ ,
\label{eq4}
\end{equation}
which shows that $\Phi$ must depend explicitly on $t$ for $E_{n} 
(x, t) \neq 0$.  There are known NLs that do not depend explicitly 
on $t$ but either only on $\dot x$ or on both $x$ and $\dot x$ 
[1,4,5,12,13,15-18]; all these NLs give $E_{n} (x) = 0$.  In the 
following we only consider the NLs that depend explicitly on $t$.  

Comparison of Eq. (\ref{eq4}) to Eq. (\ref{eq2}) shows that $E_{n} 
(x, t)$ and $L_{n} (\dot x, x, t)$ have one common term, 
and that this term becomes a NL if, and only if, $\Phi \neq \Phi (x)$ 
and  $\Phi = \Phi (t)$, which means that 
\begin{equation}
{{\partial \Phi} \over {\partial t}} = {{d \Phi} \over {d t}}\ .
\label{eq5}
\end{equation}
However, when $\Phi = \Phi (x, t)$, Eq. (\ref{eq2}) gives a
NL, but the resulting $E_{n} (x, t)$ is not a NL.  This is an 
interesting case as this extra energy term can now be added to, 
or substracted from, the standard Lagrangian $L_{s} (\dot x, x)$.  
Then, the following total Lagrangian is obtained 
\begin{equation}
L_{tot} (\dot x, x, t) = L_{s} (\dot x, x) \pm E_{n} (x, t)\ .
\label{eq6}
\end{equation}
Substitution of $L_{tot} (\dot x, x, t)$ into the EL equation gives
\begin{equation}
{{d} \over {dt}} \left ( {{\partial L_{s}} \over {\partial \dot x}} \right ) 
- {{\partial L_{s}} \over {\partial x}} = \pm {{\partial E_{n}} \over 
{\partial x}}\ . 
\label{eq7}
\end{equation}
If the standard Lagrangian is specified, the equation of motion is obtained 
from the terms that depend on $L_{s}$, and this equation represents an 
undriven (conservative) dynamical system.  However, the presence of the 
space derivative of $E_{n}$ is reponsible for converting the conservative 
equation of motion into the driven one and it can be considered to be a 
force $F(t)$.

The above results show that the addition of $E_{n}$ to $L_{s} (\dot x, x)$
modifies the original mechanical system and makes its equation of motion 
different.  The differences are of two kinds, either a time-dependent force
appears in the new equation of motion making it an inhomogeneous ODE
(driven system), or the from of the original ODE is modified by a nonlinear 
term.  In the first case, the general solutions to both the homogeneous and
inhomogeneous ODEs are the same, but the inhomogeneous ODE has also 
a complimentary solution that accounts for the force. However, in the second
case, new solutions must be found independently to each equation of motion 
with a nonlinear term added, and since the ODEs are nonlinear finding their 
solutions could be challenging [25].     

The force $F (t)$ that appears in the equation of motion is given by 
\begin{equation}
F (t) = \pm {{\partial E_{n}} \over {\partial x}} = \pm {{\partial} \over 
{\partial x}} \left ( {{\partial \Phi} \over {\partial t}} \right )\ ,
\label{eq8}
\end{equation}
which means that the initially undriven equation of motion becomes now
the driven one.  The force depends on the form of the gauge function, 
$\Phi (x, t)$, which must be an explicit function of both the dependent variable 
and the independent variable.  Since the gauge function $\Phi (x, t)$
can be any differentiable scalar function, the resulting force can also 
be of any form (see Section 3 for specific applications). 

According to Eq. (\ref{eq8}), the simplest form of the gauge function 
$\Phi (x, t)$ that gives a non-zero force is $\Phi_1 (x, t) = c_1 x t$, 
where $c_1$ is an arbitrary real constant.

Thus, the resulting force is $F_1 = \pm c_1$ = const [16-18].  In 
the following, we generalize this result to gauge functions that 
depend on arbitrary functions of $x$ and $t$ but keep the variables 
separated to obtain general analytical results.  

\subsection{Gauge functions and forces}
  
Let us generalize the gauge function $\Phi_1 (x, t) = c_1 x t$ and 
consider  
\begin{equation}
\Phi_{g1} (x, t) = \sum_{m=1}  \sum_{n=1} C_{m,n} x^{m} 
t^{n}\ ,
\label{eq9}
\end{equation}
where $C_{m,n}$ are arbitrary real constants, and $m$ and $n$ are 
positive integers.  Since, for dynamical systems, the variables $t$ and 
$x$ represent time and displacement, respectively, both variables have 
dimensions. It is therefore required that the constants $C_{m,n}$ have 
different dimensions to guarantee that each term in the expansion has 
the same physical units as the gauge function [$kg\ m^2 / s$].  Thus, 
the dimensions of $C_{1,1} = c_1$ are [$kg\ m / s^2$], which are of the 
dimensions of force (see example at the end of Section 2.2).

The general gauge function $\Phi_{g1} (x, t)$ gives the following NL
\begin{equation}
L_{ng1} (\dot x, x, t) = \sum_{m=1}  \sum_{n=1} C_{m,n} 
\left [ m \dot x t + n x \right ] x^{m-1} t^{n-1}\ ,
\label{eq10}
\end{equation}
where each term of this summation is also a NL.  The general force 
resulting from $\Phi_{g1} (x, t)$ is
\begin{equation}
F_{g1} (x, t) = \sum_{m=1}  \sum_{n=1} m n\ C_{m,n}\ x^{m-1} 
t^{n-1}\ .
\label{eq11}
\end{equation}
In special case of $m = n = 1$, the general force $F_{g1} (x, t)$ reduces
to the force $F_{1} = C_{1,1} = c_1$ = const.  The resulting general 
force is a power series in both the dependent and independent variables.

To further generalize the gauge function $\Phi_{g1} (x, t)$, we consider
\begin{equation}
\Phi_{g2} (x, t) = \sum_{m=1} c_{m} x^{m} f_m (t)\ ,
\label{eq12}
\end{equation}
where $c_{m}$ are arbitrary constants of real values and different dimensions, 
and $m$ and $n$ are positive integers.  In addition, $f_m (t)$ are arbitrary 
functions that are required to be ordinary ($f_m :\ \mathcal{R} \rightarrow 
\mathcal{R}$) and smooth ($\mathcal{C}^{\infty}$).  The resulting general 
NL is 
\begin{equation}
L_{ng2} (\dot x, x, t) = \sum_{m=1} \left [ m f_{m} (t) \dot{x} (t) 
+ \dot f_m (t) x \right ] c_m x^{m-1}\ ,
\label{eq13}
\end{equation}
and the general force is
\begin{equation}
F_{g2} (x, t) = \sum_{m=1} m \dot f_m (t) c_m x^{m-1}\ .
\label{eq14}
\end{equation}
The resulting force is a power series in the dependent variable
with each term of this power being multiplied by any differentiable 
function of the independent variable. This general force reduces to 
$F_{g2} = F_{1} = c_1$ = const if, and only if, $m = 1$ and 
$f_1 (t) = t$; however, for different forms of $f_1$,
different 
forms of NLs are obtained.  

Finally, we generalize $\Phi_{g2} (x, t)$ to   
\begin{equation}
\Phi_{g3} (x, t) = \sum_{m=1} \sum_{n=1} c_{m,n} f_m (t) 
g_n (x)\ ,
\label{eq15}
\end{equation}
where $g_n (x)$ are arbitrary functions that are required to be ordinary 
($g_n :\ \mathcal{R} \rightarrow \mathcal{R}$) and smooth 
($\mathcal{C}^{\infty}$).   The gauge function gives the following 
general NL 
\begin{equation}
L_{ng3} (\dot x, x, t) = \sum_{m=1} \sum_{n=1} c_{m,n} 
\left [ \dot f_m (t) g_n (x) + \dot{x} (t) f_{m} (t) g_n^{\prime} (x) \right ]\ ,
\label{eq16}
\end{equation}
and the general force
\begin{equation}
F_{g3} (x, t) = \sum_{m=1} c_{m,n} \dot f_m (t) g_n^{\prime} (x)\ ,
\label{eq17}
\end{equation}
where $g_n^{\prime} (x)$ is the derivative of $g_n (x)$ with respect 
to $x$.  The gauge function $\Phi_{g3} (x, t)$ is the most general one 
that can be considered when the dependent and independent variables 
are separated.  Therefore, the resulting NL $L_{ng3} (\dot x, 
x, t)$ and the force $F_{g3} (x, t)$ are also the most general that can
be obtained under these conditions, and they are given as infinite sums
of all differentiable functions.  It must be pointed out that each term in
the power series given by Eqs. (\ref{eq11}) and (\ref{eq14}) and in 
the summation of functions given by Eq. (\ref{eq17}) represents a 
partial NL.

The functions $f_m (t)$ and $g_n (x)$ can be any known elementary 
functions, including algebraic, trigonometric, exponential, logarithmic, 
hyperbolic, inverse trigonometric and hyperbolic, and others.  The 
summation of the functions $f_m (t)$ and $g_n (x)$ gives a significant 
amount of flexibility in defining different forces by using the elementary 
functions because each term in the summation is a NL that leads to a 
non-zero force.  Since $f_m (t)$ and $g_n (x)$ are to be specified, 
most known forces in classical mechanics can be formally introduced 
this way.  The coefficients $c_{m,n}$ can be determined by the force 
required for a given physical system; for instance, the coefficients may 
represent the amplitude of the force for the given system. 

An interesting result is that the above method to define forces can also 
be extended to introduce nonlinearities into otherwise linear dynamical 
systems.  This shows universality and a broad range of applications of 
the presented results to classical dynamics.

\section{Physical applications to classical dynamics}

\subsection{Forces in dynamical systems}

Our theoretical results presented in the previous section demonstrate
how NLs and their gauge functions can be used to introduce forces 
to classical dynamical systems.  We consider the equation of motion 
for a driven harmonic oscillator given by 
\begin{equation}
\ddot{x} (t) + x (t) = F (t)\ ,
\label{eq18}
\end{equation}
where typical forms of the force $F (t)$ are given in Table I.  The
presented forces are time-dependent and they are given by some 
well-known elementary functions.  

Our aim is to identify gauge functions that can be used to introduce 
these forces.  For the forces presented in Table 1, the most appropriate 
is the general gauge function given by Eq. (\ref{eq15}).  The procedure 
is straightforward and requires comparing the forces in Table 1 to the 
general force term given by Eq. (\ref{eq17}). In other words, we identify 
the coefficients $c_{m,n}$ and the derivatives of functions $f_m (t)$ 
and $g_n (x)$.  Then, Eq. (\ref{eq15}) can be used to determine the 
gauge function corresponding to each force of Table 1 as well as the 
NL corresponding to each force, which can be obtained 
using Eq. (\ref{eq16}).

We may also consider $F$ that does not explicitly depend on $t$ but instead
depends on $x$.  Taking $F(x) = - \varepsilon F_0 x$, where $0 < \varepsilon
\leq 1$, and substituting into Eq. (\ref{eq17}), we obtain the equation of motion
of the altered simple harmonic oscillator [24].  The gauge function for this force
is $\Phi= -\frac{\varepsilon}{2} F_0 x^2 t$ as obtained from Eq. (\ref{eq15}), 
and the corresponding NL is $L_n (\dot x, x, t) = - \frac{1}{2}\varepsilon F_0 
(2 \dot {x} t + x) x $.
%
%
\begin{center}
\centering                                          
\begin{tabular}{|c|c|c|}                      
\hline                                      
Dynamical & Force & Gauge function \\
systems & $F (t)$  & $\Phi(x,t) = \Phi_1(x,t) + \Phi_2(x,t)$ \\
\hline
Driven                & $F(t)=F_0 \cos  t $  &  $\Phi_1 (x,t)= x F_0  \sin t$ \\
oscillators            &                              &  $\Phi_2 (x,t)= 0$ \\ \cline{2-3} \\

                         & $F(t)=F_0 \cos^2 t$ & $\Phi_1 (x,t)= \frac{1}{2} x t F_0$ \\  
                         &  $= \frac{1}{2} F_0 (1+\cos 2t)$ & $\Phi_2 (x,t)= \frac{1}{4} 
                             x F_0 \sin 2  t$ \\ \cline{2-3} \\ 
  
                         & $F(t)=F_0 \cos^3 t$ & $\Phi_1 (x,t) =  \frac{3}{4} x F_0 \sin t$ \\  
                         &  $= \frac{1}{4} F_0 (3\cos t + \cos 3t)$  & $\Phi_2(x,t) =  \frac{1}{12} 
                             x F_0 \sin 3 t$ \\ \cline{2-3}\\     
          
                         & $F(t)= F_0 e^{i t}$  & $\Phi_1 (x,t)= -ix F_0 e^{i t}$ \\  
                         &                               &  $\Phi_2 (x,t)= 0$  \\ \cline{2-3} \\   
       
                         & $F(t)=F_1 \cos t$     & $\Phi_1(x,t)= x F_1 \sin t$  \\   
                         &    $+ F_2 \sin t$       & $\Phi_2(x,t)= - x F_2 \cos t$  \\    
\hline            
RLC circuits        & $E(t)=E_0 \sin t$ & $\Phi_1(x,t)=-x E_0 cos t$ \\
                        &                           & $\Phi_2(x,t)=0$ \\
\hline                
\end{tabular}
\end{center}
Table 1. Selected forces in dynamical systems [22,23,25] and their gauge functions.
%

The gauge functions presented in Table 1 become special cases of general gauge 
functions given by Eqs (\ref{eq9}), (\ref{eq12}) and (\ref{eq15}).  As an example, 
let us consider the force $F (t) = \cos^2 t = \frac{1}{2} F_0 (1+\cos 2t)$.  Then,
its gauge function $\Phi_1 (x,t)= \frac{1}{2} x t F_0 = \Phi_{g1} (x,t)$ with 
$m = n = 1$ and $C_{1,1} = \frac{1}{2} F_0$ (see Eq. \ref{eq9}).  However, 
the gauge function $\Phi_2 (x,t)= \frac{1}{4} x F_0 \sin 2t = \Phi_{g2} (x,t)$ 
with $m = 1$ and $C_{1} = \frac{1}{4} F_0$ and $f_1(t) = \sin 2t$ (see Eq. 
\ref{eq12}).  The same force can also be identified as $\Phi_{g1} (x,t) = C_{1,1}
f_1 (t) g_1 (x) + C_{2,2} f_2 (t) g_2 (x)$, where $C_{1,1} = \frac{1}{2} F_0$,
$f_1 (t) = t$, $g_1 (x) = x$, $C_{2,2} = \frac{1}{4} F_0$, $f_2 (t) = \sin 2t$ 
and $g_2 (x) = x$.  Similar identification can be performed for all forces given 
in Table 1.

In the above results, the gauge functions were obtained for forces that are 
typically used to drive harmonic oscillators.  However, the process can be
reversed and forces can be determined by using the gauge functions 
presented in Section 2.  As we show in the following, the same method 
can also be used to introduce nonlinearities to the equation of motion of 
a harmonic oscillator.

\subsection{Nonlinearities in dynamical systems}

After demonstrating how forces can be added to the equation
of motion of a harmonic oscillator, we now show that our method also allows
introducing nonlinearities to this equation, and therefore converting the equation
into one of the well-known nonlinear equations of motion in classical dynamics.

We write the following equation of motion 
\begin{equation}
\ddot{x} (t) + x (t) = H (\dot x, x)\ ,
\label{eq20}
\end{equation}
where $H (\dot x, x)$ can represent different forms of nonlinearities in dynamical 
systems.  We have selected several examples of nonlinear dynamical systems, 
including well-known systems such as the Duffing oscillator [23,25], and present 
them in Table 2.

The presented method of introducing nonlinearities into otherwise linear equations 
of motion can be combined together with the method of defining forces, so that linear, 
undriven dynamical systems can be converted into nonlinear and driven ones.  
%
%
\begin{center}
\centering                                          
\begin{tabular}{|c| c| c|}                      
\hline                                      
Type of  & Nonlinearity & Gauge function\\ 
oscillator & $H (x)$  & $\Phi(x,t)$ \\ [0.5ex] 
\hline
Quadratic   &	$H (x) = - \varepsilon x^2$  &  $\Phi (x,t) =- \frac{1}{3}\varepsilon x^3t$ \\ 
Duffing  	    &	$H (x) = - \varepsilon x^3$  &  $\Phi (x,t) =-\frac{1}{4}\varepsilon x^4t$ \\ 
Quadratic and cubic  &  $H (x) = - \varepsilon (x^2 + x^3) $  &  $\Phi (x,t)=- \frac{\varepsilon}{3} 
x^3t-\frac{\varepsilon}{4} x^4t$  \\ 
Quartic    &	$H (x) = - \varepsilon x^4$  &  $ \Phi (x,t) =-\frac{1}{5}\varepsilon x^5t$ \\ 
Quintic    &	$H (x) = - \varepsilon x^5$  &  $ \Phi (x,t) =-\frac{1}{6}\varepsilon x^6t$  \\ 
Higher-order   &	$H (x) = - \varepsilon x^{2n+1}$  & $ \Phi (x,t)=-\frac{\varepsilon}{2n+2}  
x^{2n+2}t  $        \\ 
\hline                
\end{tabular}
\end{center}
Table 2. Selected nonlinearities in dynamical systems [22-24] and their gauge functions.

The gauge functions presented in Table 2 become special cases of general gauge 
functions given by Eqs (\ref{eq9}), (\ref{eq12}) and (\ref{eq15}).  It is easy to 
show that any $\Phi (x,t)$ of Table 2 can be identified as either $\Phi_{g1} (x,t)$,
$\Phi_{g2} (x,t)$ or $\Phi_{g3} (x,t)$ by making appropriate selection of the 
coefficients and functions in these equations.  It must also be pointed out that 
more general nonlinearities can be obtained from the results of Section 2 than
those presented in Table 2.   

All dynamical systems considered in Section 3 illustrate the relationships between forces
and nonlinearities and their corresponding gauge functions.  The purpose of this illustration 
is to demonstrate that typical forms of forces and nonlinearities have gauge functions that 
can be used to introduce them by way of the method presented in this paper.  Moreover, 
our results also show that one may consider a linear and undriven equation of motion 
and specify gauge functions to convert this equation into a nonlinear and driven equation, 
which means that our method can be applied to a broad range of dynamical systems.

\subsection{Comparison to previous results}

For each gauge function given in Tables 1 and 2, the corresponding NL can 
be obtained by taking the total derivative of this function with respect to time.  
Recently, it was shown [26] that if the total derivative of the null Lagrangian 
$L_{null}$, generated by these gauge functions, is zero, then every NL gives 
an equation of motion; note that here, $L_{null}$is either $L_{ng1}$ or 
$L_{ng2}$ or $L_{ng3}$ derived in Section 2.3.  Moreover, the above 
condition requires that the inverse of any NL becomes a special non-standard 
Lagrangian [NSL], which means that there is a corresponding NSL for each 
derived NL [26].  In other words, for each gauge function listed in Tables 1 
and 2, there is a corresponding NSL, which requires to be compared to those 
already known from the literature. There are several different families of NSLs 
that have been considered by different authors and in the following we briefly 
summarize the previous work and compare their results to those obtained in 
this paper.  

The original form of NSL introduced by Arnold [27] has been used in a 
number of papers and applied to a variety fo dynamical systems [28-32].  
The forms of the NSLs considered in these papers are very similar to those 
resulting from the inverse of the derived NLs.  However, the NSLs obtained 
here are more restricted than those found before because of the null condition 
that all NLs must obey [26].  This null condition requires that the NLs obey 
some relationships between the coefficients of the equations of motion [26]; 
no such relationships limit the NSLs derived in the previous papers.  Thus, 
the previously found NSLs are more general than those obtained in this paper,
which also means that the previous NSLs do not have corresponding NLs.

A distinct family of NSLs was introduced by El-Nablusi [33-38], who also 
applied them to a variety of physics and astronomy problems.  The NSLs are 
different than the NSLs presented in this paper and so far no studies have 
been done regarding the existence of corresponding NLs for them.  Similarly, 
the NSLs calculated by Nucci and Leach [39,40] and others [41] are found by 
using the Jacobi Last Multiplier method; again, these are of very different 
forms than those found here.  Another family of NSLs was introduced by 
Havas [42], see also [42,43], and they are specifically constructed for harmonic 
oscillators.  It has not been yet been determined whether the NSLs studied 
in [39-43] have their underlying NLs or not; this is a project for future 
investigations.

Finally, it must be pointed out that the NLs can be also divided into two 
families, namely, standard and non-standard NLs, and that all NLs derived 
and studies in this paper are the former; the existence of the latter has been 
demonstrated in [44] and their relationships to the standard NLs obtained 
here remains to be investigated.

\section{Conclusion}

A family of general NLs and their gauge functions were constructed for second-order 
ordinary differential equations of motion in one dimension, and it was show that, in 
special cases, the derived NLs reduce to those already known.  The main advantage 
of the constructed NLs and gauge functions is that they may be used to define 
forces. The presented results significantly extend the previous work in this area 
and demonstrate how to convert undriven dynamical systems into driven ones. 
An interesting result is that the presented method can also be used to formally 
convert a linear dynamical system into a nonlinear one.  

The most important conclusion of our work is that forces and nonlinearities 
known in classical dynamics have corresponding gauge functions, which 
means that, by the specification of gauge functions, many linear and 
undriven dynamical systems may be formally converted into nonlinear 
and driven systems.  Thus, the presented results demonstrate a new 
important role of gauge functions and null Lagrangians in classical 
mechanics and, specifically, in its theory of dynamical systems.

{\bf Acknowledgments}
We thank B. D. Tran for the verification of our results using 
{\it Mathematica}, and for his useful comments on an earlier 
version of this paper.


\section{References}

\begin{thebibliography}{999}

\bibitem{1} P.J. Olver, Applications of Lie Groups to Differential Equations, 
                 Springer-Verlag, New York, 1993

\bibitem{2} H. Helmholtz, On physical meaning of the principle of least action, J. f. d. 
                  Reine u. Angew Math., 100 (1887) 213

\bibitem{3} J. Lopuszanski, The Inverse Variational Problems in Classical Mechanics, 
                  World Scientific, Singapore, 1999

\bibitem{4} P.J. Olver and J. Sivaloganathan, The structure of null Lagrangians, 
                  Nonlinearity 1 (1989) 389

\bibitem{5} M. Crampin and D.J. Saunders, On null Lagrangians, Diff. Geom. and its Appl., 
                   22, 131, 2005 

\bibitem{6} R. Vitolo, On different geometric formulations of Lagrange formalism, Diff. Geom. 
                   and its Appl. 10 (1999) 293

\bibitem{7} D.E. Betounes, J. Math. Phys. 28 (1987) 2347 

\bibitem{8} D. Krupka. O. Krupkova, and D. Saunders, The Cartan form and its generalizations 
                    in the calculus of variations, Int. J. Geom. Meth. Mod. Phys. 7 (2010) 631 

\bibitem{9} D.R. Grigore, Trivial second-order Lagrangians in classical field theory,  
                  J. Phys. A 28 (1995) 2921

\bibitem{10} D. Krupka and J. Musilova, Trivial Lagrangians in field theory, Diff. Geom. and its 
                   Appl. 9 (1998) 225

\bibitem{11} M. Giaquinta and S. Hilderbrandt, Calculus of Variations I, Springer, Berlin, 1996.

\bibitem{12} D.R. Anderson, D.E. Carlson and E. Fried, A continuum-mechanical theory for 
                   nematic elastomers, J. Elasticity 56 (1999) 35

\bibitem{13} G. Saccomandi and R. Vitolo, R., Null Lagrangians for nematic elastomers, J. Math. 
                    Sci. 136 (2006) 4470

\bibitem{14} J.-M. Levy-Leblond, Group theoretical foundations of classical mechanics: the 
                   Lagrange gauge problem, Comm. Math. Phys. 12 (1969) 64

\bibitem{15} Z. E. Musielak and T. B. Watson, Gauge functions and Galilean invariance of 
                   Lagrangians, Phys. Let. A 384 (2020) 126642

\bibitem{16} Z.E. Musielak, T.B. Watson, General null Lagrangians, exact gauge functions 
                   and forces in Newtonian mechanics, Phys. Let. A, 384 (2020) 126838

\bibitem{17} Z.E. Musielak, L.C. Vestal, B.D. Tran, T.B. Watson, Gauge functions in 
                   classical mechanics: From undriven to driven dynamical systems, Physics, 
                   2 (2020) 425

\bibitem{18} L.C. Vestal, Z.E. Musielak, Bateman Oscillators: Caldirola-Kanai and null 
                    Lagrangians and gauge functions, Physics, 3 (2021) 449

\bibitem{19} H. Bateman, On dissipative systems and related variational 
                   principles, Phys. Rev. 38 (1931) 815

\bibitem{20} P. Caldirola, Forze non conservative nella meccanica quantista, Nuovo Cim., 
                   18 (1941) 393.

\bibitem{21} E. Kanai, On the quantization of the dissipative systems, Prog. Theor. Phys., 
                   3 (1948) 44.

\bibitem{22} H. Goldstein, C.P. Poole and J.L. Safko, Classical Mechanics (3rd Edition), 
                   Addison-Wesley, San Francisco, CA, 2002  

\bibitem{23} J.V. Jos\'e and E.J. Saletan, Classical Dynamics, A Contemporary Approach,
                   Cambridge Univ. Press, Cambridge, 2002

\bibitem{24} R. Abraham and J.E. Marsden, Foundation of Mechnanics, AMS Chelsea
                    Publishing, Prividence, RI, 2008 

\bibitem{25} P.B. Kahn, Mathematical Methods for Scientists and Engineers, Wiley \& Sons, 
                    New York, 1990

\bibitem{26} R. Das and Z.E. Musielak, General Null Lagranians and Their Novel Role 
                    in Classical Dynamics, Phys. Scripta, 2022, submitted; 
                    arXiv:2203.04470v2 [math-ph] 24 March 2022

\bibitem{27} V.I. Arnold, Mathematical Methods of Classical Mechanics, Springer, New York, 
                   1978 

\bibitem{28} Z.E. Musielak, Standard and non-standard Lagrangians for dissipative 
                   dynamical systems with variable coefficients, J. Phys. A Math. Theor. 
                   41 (2008) 055205.

\bibitem{29} Z.E. Musielak, General conditions for the existence of non-standard Lagrangians 
                   for dissipative dynamical systems, Chaos, Solitons  Fractals 42 (2009) 2640.

\bibitem{30} J.L. Cie\'sli\'nski and T. Nikiciuk, A direct approach to the construction of 
                   standard and non-standard Lagrangians for dissipative-like dynamical systems 
                   with variable coefficients, J. Phys. A Math. Theor. 43 (2010) 175205. 

\bibitem{31} A. Saha and B. Talukdar, Inverse variational problem for nonstandard Lagrangians,
                   Rep. Math. Phys. 73 (2014) 299--309. 

\bibitem{32} N. Davachi and Z.E. Musielak, Generalized non-standard Lagrangians, 
                   J. Undergrad. Rep. Phys. 29 (2019) 100004.

\bibitem{33} R.A. El-Nabulsi, Nonlinear dynamics with non-standard Lagrangians, 
                   Qual. Theory Dyn. Syst. 13 (2013) 273.

\bibitem{34} R.A. El-Nabulsi, A generalized nonlinear oscillator from non-standard degenerate 
                   Lagrangians and its consequent Hamiltonian formalism, Proc. Natl. Acad. Sci., India, 
                    Sect. A Phys. Sci. 84 (2014) 563.

\bibitem{35} R.A. El-Nabulsi, Fractional oscillators from non-standard Lagrangians and time-dependent 
                   fractional exponent, Comp. Appl. Math. 33 (2014) 163.

\bibitem{36} R.A. El-Nabulsi, New non-standard Lagrangians for the Liénard-type equations, Appl. 
                   Math. Lett. 63 (2017) 124.

\bibitem{37} R.A. El-Nabulsi and W. Anukool, A new approach to nonlinear quartic oscillators, 
                   Arch. Appl. Mech. 92 (2022) 351.

\bibitem{38} N.A. Kudryashov, New non-standard Lagrangians for the Liénard-type oscillator, 
                   Appl. Math. Lett. Phys. 63 (2017) 124.

\bibitem{39} M.C. Nucci and P.G.L. Leach, Lagrangians galore, J. Math. Phys. 48 (2007) 123510. 

\bibitem{40} M.C. Nucci and P.G.L. Leach, Jacobi Last Multiplier and Its Applications in Mechanics, 
                    Phys. Scripta, 78, 065011, 2008
      
\bibitem{41} A.G. Choudhury, P. Guha and B. Khanra, On the Jacobi last multiplier, 
                    integrating factors and the Lagrangian formulation of differential equations 
                    of the Painlevé–Gambier classification, J. Math.Anal. Appl. 360 (2009) 651.

\bibitem{42} P. Havas, The Range of Application of the Lagrange Formalism, Nuovo Cim. 5 (1957) 363. 

\bibitem{43} G. Gonzalez, Comment on "Standard and non-standard Lagrangians for dissipative 
                   dynamical systems with variable coefficients'; 
                   arXiv:2202.05391v1 [physics.class-ph] 3 Feb 2022

\bibitem{44} Z.E. Musielak, Nonstandard Null Lagrangians and Gauge Functions for Newtonian 
                     Law of Inertia, Phys. 3 (2021) 903.


\end{thebibliography}
\end{document}